\documentclass{elsart_mpi}
\usepackage{epsfig}
\usepackage{rotating}

\begin{document}

\begin{frontmatter}

\title{Construction and Test of MDT Chambers for the
ATLAS Muon Spectrometer}

\author{F.~Bauer, U.~Bratzler, H.~Dietl},
\author{H.~Kroha\thanksref{email2}},\author{Th.~Lagouri},
\author{A.~Manz, A.~Ostapchuk, R.~Richter, S.~Schael}
\address{Max-Planck-Institut f\"ur Physik, F\"ohringer Ring 6, D-80805 Munich, Germany}
\author{S.~Chouridou, M.~Deile, O.~Kortner, A.~Staude,}
\author{R.~Str\"ohmer and T.~Trefzger}
\address{Ludwig-Maximilians-Universit\"at, Schellingstra\ss e 4, D-80799 Munich, Germany}
\thanks[email2]{Corresponding author, e-mail: kroha@mppmu.mpg.de}

\begin{abstract}
The Monitored Drift Tube (MDT) chambers for the muon spectrometer
of the ATLAS detector at the Large Hadron Collider (LHC) consist of
3--4 layers of pressurized drift tubes on either side of
a space frame carrying an optical monitoring system to correct for
deformations. The full-scale prototype of a large
MDT chamber has been constructed with methods suitable for large-scale production.
X-ray measurements at CERN showed a positioning accuracy of the sense wires
in the chamber of better than the required $20~\mu$m (rms).
The performance of the chamber was studied in a muon beam at CERN.
Chamber production for ATLAS now has started.
\end{abstract}

\end{frontmatter}

The ATLAS muon spectrometer~\cite{TDR} is designed to provide stand-alone
muon momentum resolution of $\Delta p_T/p_T = 2-10~\%$
for transverse momenta between 6~GeV and 1~TeV over a pseudo-rapidity
range of $|\eta |\le 2.7$.  
This requires very accurate track sagitta measurement with three layers of
muon chambers in a superconducting air-core toroid magnet
and high-precision optical monitoring systems to correct 
for chamber misalignment. Precision drift chambers,
the Monitored Drift Tube (MDT) chambers, have been developed to provide
a track position resolution of $40~\mu$m over an active area of
5500~m${}^2$.
\begin{figure}[h]
\vspace{-5mm}
\begin{tabular}{ll}
\begin{minipage}{0.4\textwidth}
\hspace{-5mm}
\includegraphics[width=1.10\textwidth]{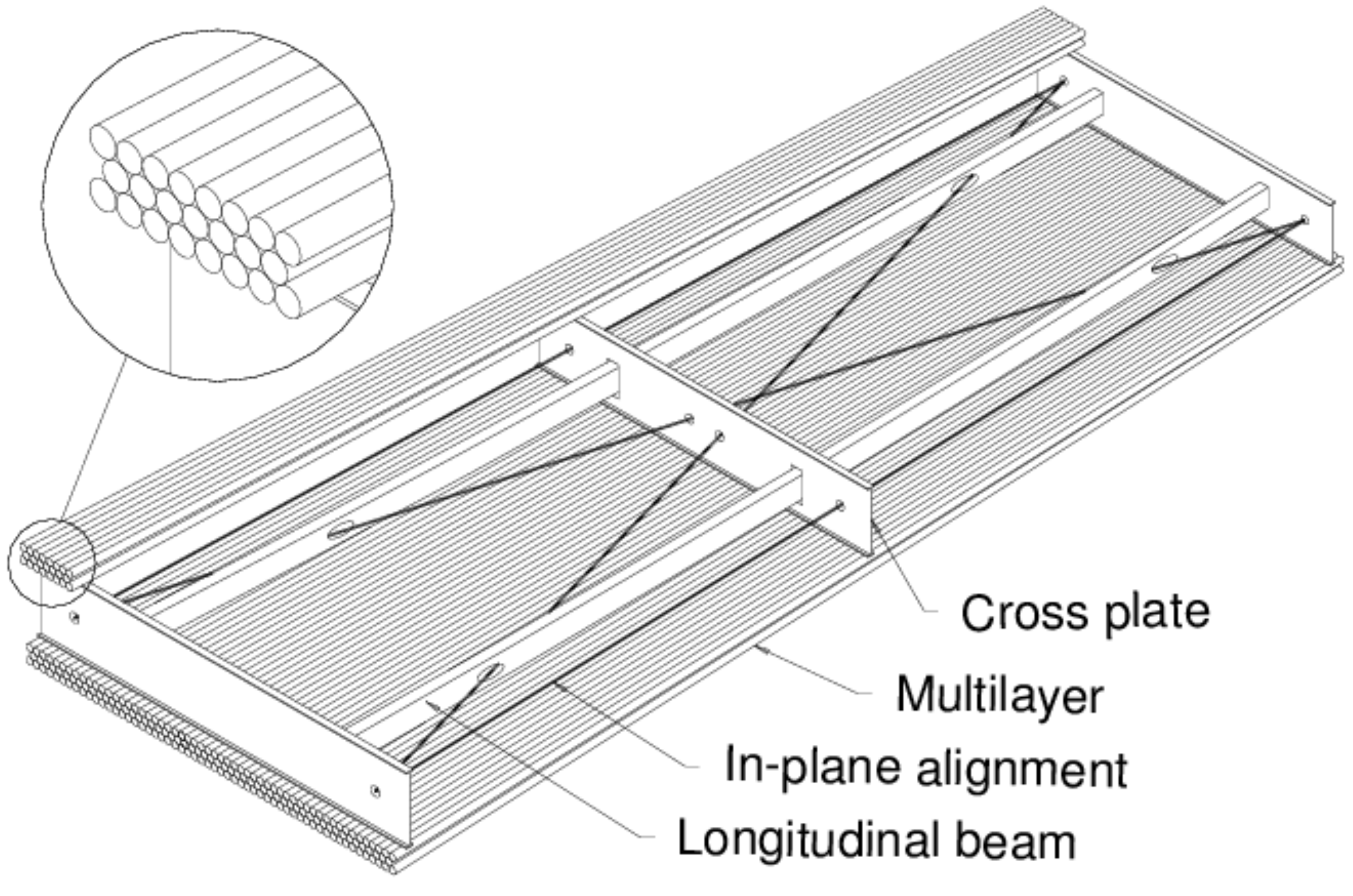}
\vspace{10mm}
\caption{MDT chamber for the\hfill\break
ATLAS muon spectrometer.}\label{MDT}
\end{minipage} &
\begin{minipage}{0.6\textwidth}
\includegraphics[angle=-90,width=0.96\textwidth]{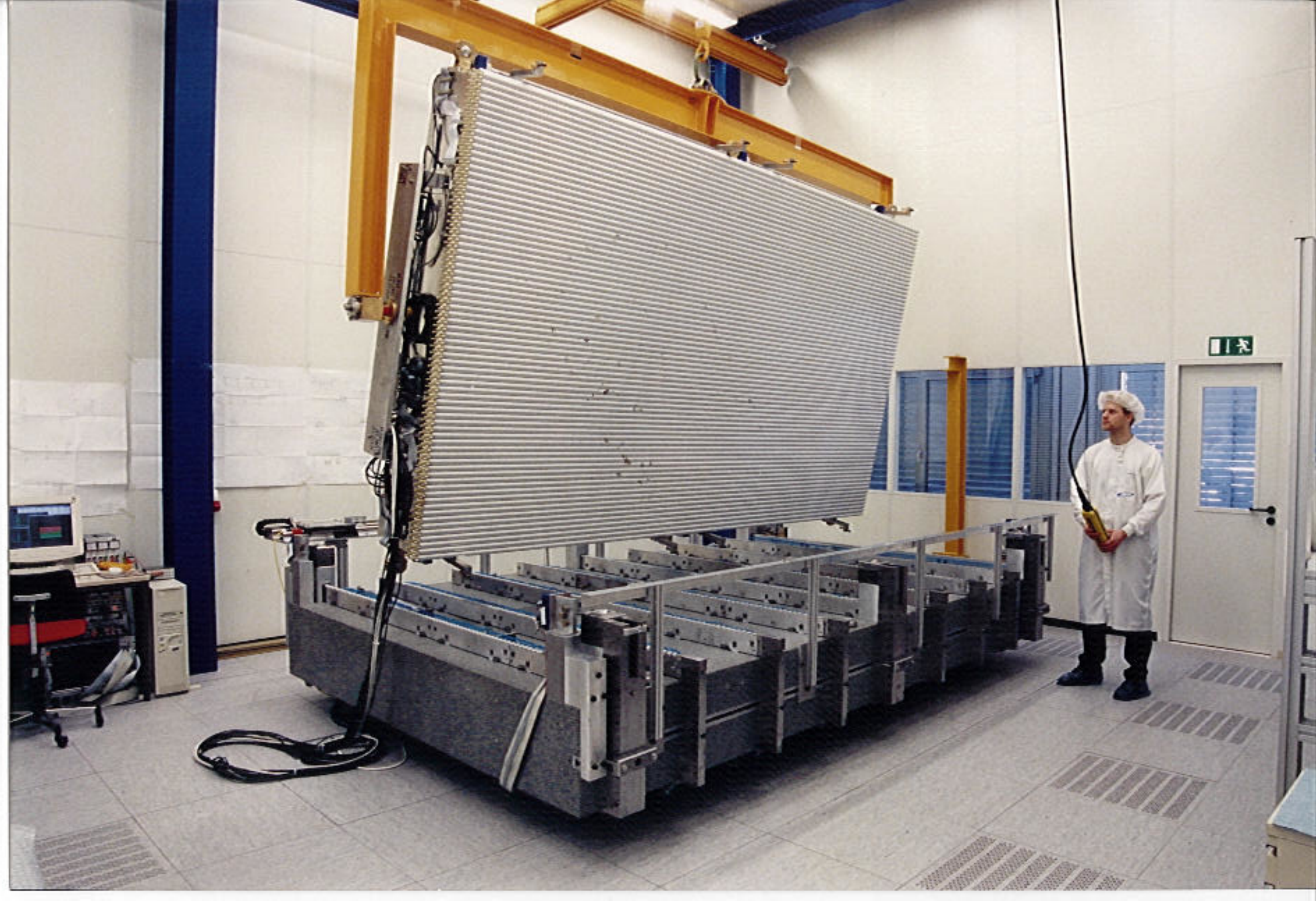}

\vspace{1mm}
\caption{The MDT prototype chamber\hfill\break
 during assembly.}\label{BOS}
\end{minipage} \\
\end{tabular}
\end{figure}

The MDT chambers (see Fig.~\ref{MDT}) consist of 3--4 layers of precision
aluminum drift tubes with $29.970\pm 0.015$~mm outer diameter and $400~\mu$m wall thickness
on either side of a space frame carrying an optical monitoring system to 
correct for chamber deformations. The drift tubes are operated at a gas pressure of 3~bar 
to provide a single-tube position resolution of at least $80~\mu$m (rms)
at the low gas gain of $2\times 10^4$ used to minimize ageing effects.
The sense wires of the drift tubes have to be positioned
in the chamber with an accuracy of $20~\mu$m (rms) for a chamber position resolution of 
$40~\mu$m (rms).
\begin{figure}[h]
\begin{tabular}{ll}
\begin{minipage}{0.53\textwidth}
\hspace{-3mm}
\includegraphics[width=1.00\textwidth]{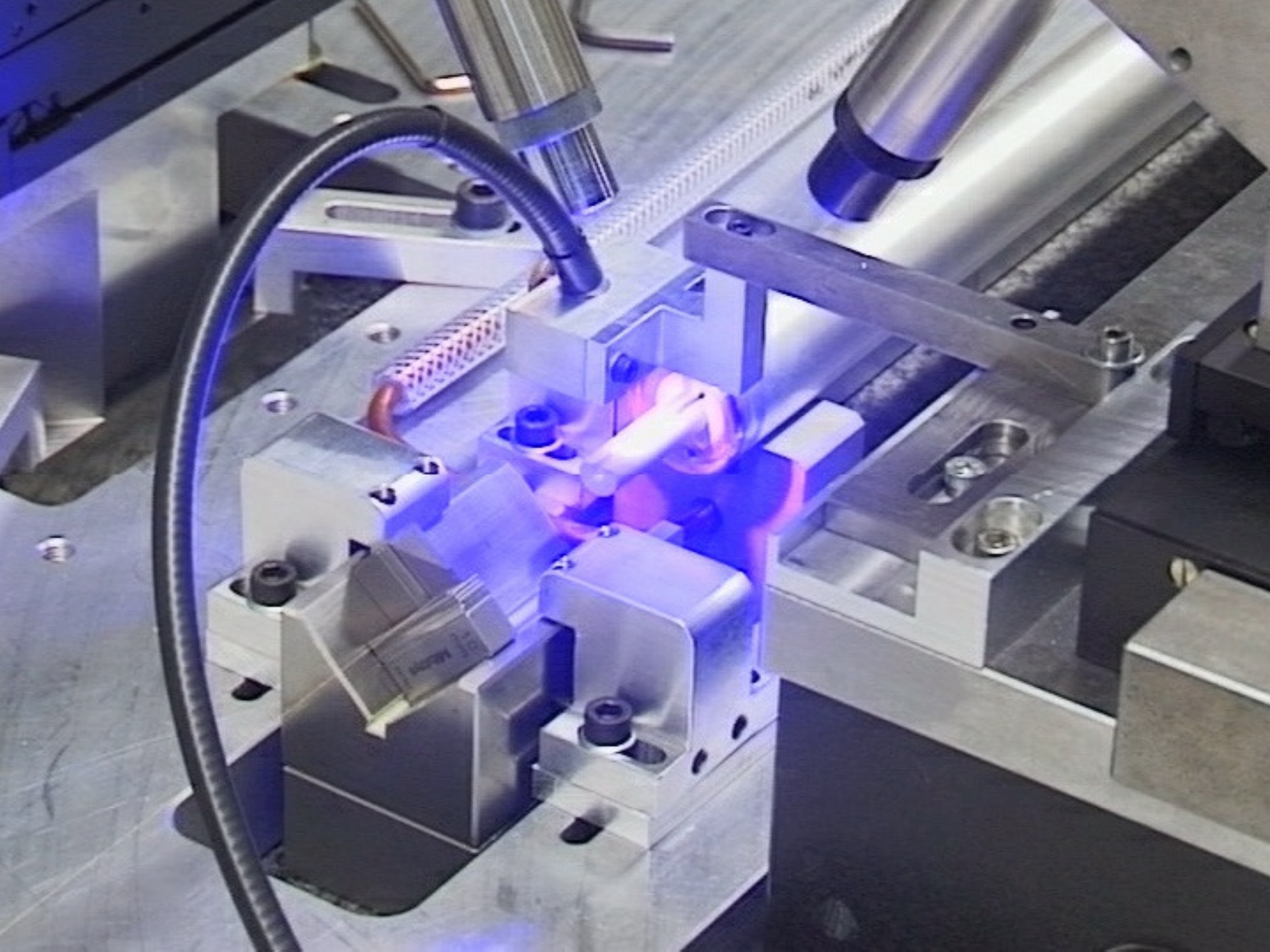}
\caption{Precise wire positioning using\hfill\break
glue curing under UV illumination.}\label{glue}
\end{minipage} &
\begin{minipage}{0.47\textwidth}
\hspace{-12mm}
\includegraphics[width=1.20\textwidth]{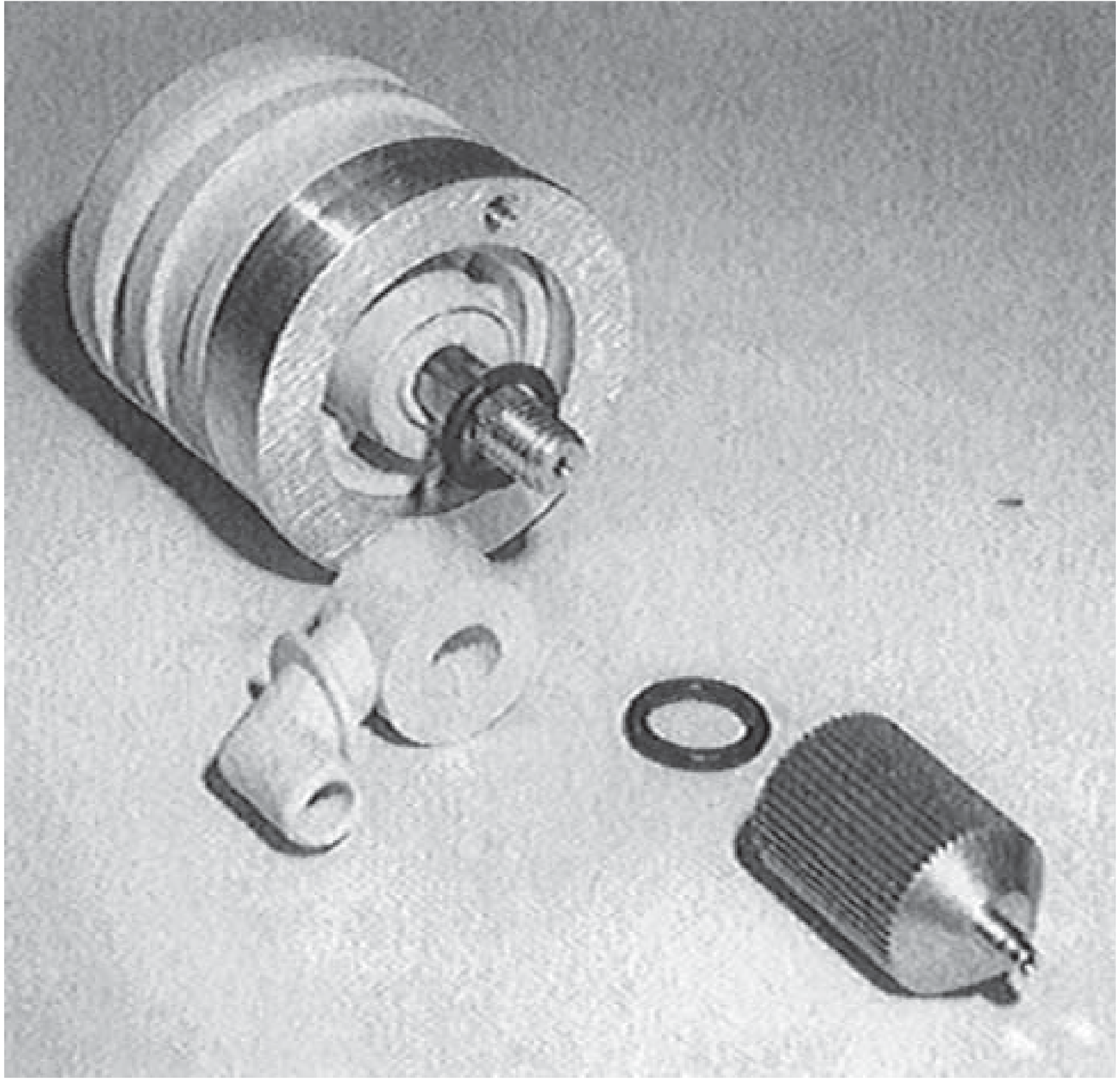}

\caption{Endplug design for the\hfill\break
wire glueing technique.}\label{uvplug}
\end{minipage} \\
\end{tabular}
\end{figure}
 
For the ATLAS muon spectrometer,
1200 MDT chambers containing about 400000 tubes of 1-6~m length have to be built
at 12 construction sites over a period of 4 years.
With the construction of a full-scale prototype chamber with 432 tubes of 3.8~m
length and a width of 2~m in 1998 (see Fig.~\ref{BOS})
it has been demonstrated
that the high mechanical precision can be achieved with assembly methods~\cite{TDR}
suitable for large-scale production.

The construction of a MDT chamber consists of two major steps,
the fabrication of the drift tubes with accurately positioned sense wires
and the precise assembly of the tubes in the chamber.
For the prototype chamber, a drift tube fabrication method was used which
allowed to position the wires in the precise aluminum tubes with an accuracy
of $\sigma\approx 10~\mu$m in each coordinate
by means of external references and fast-curing glue (see Fig.~\ref{glue}). 
The glue has been shown to cause no
ageing of the drift tubes with Ar:CO${}_2$ gas mixtures.
The wire locations in the tubes have been measured with stereo microscopes
during assembly and with an X-ray method after assembly.
\begin{figure}[h]
\vspace{1mm}
\begin{tabular}{ll}
\begin{minipage}{0.52\textwidth}
\hspace{-8mm}
\includegraphics[width=1.14\textwidth]{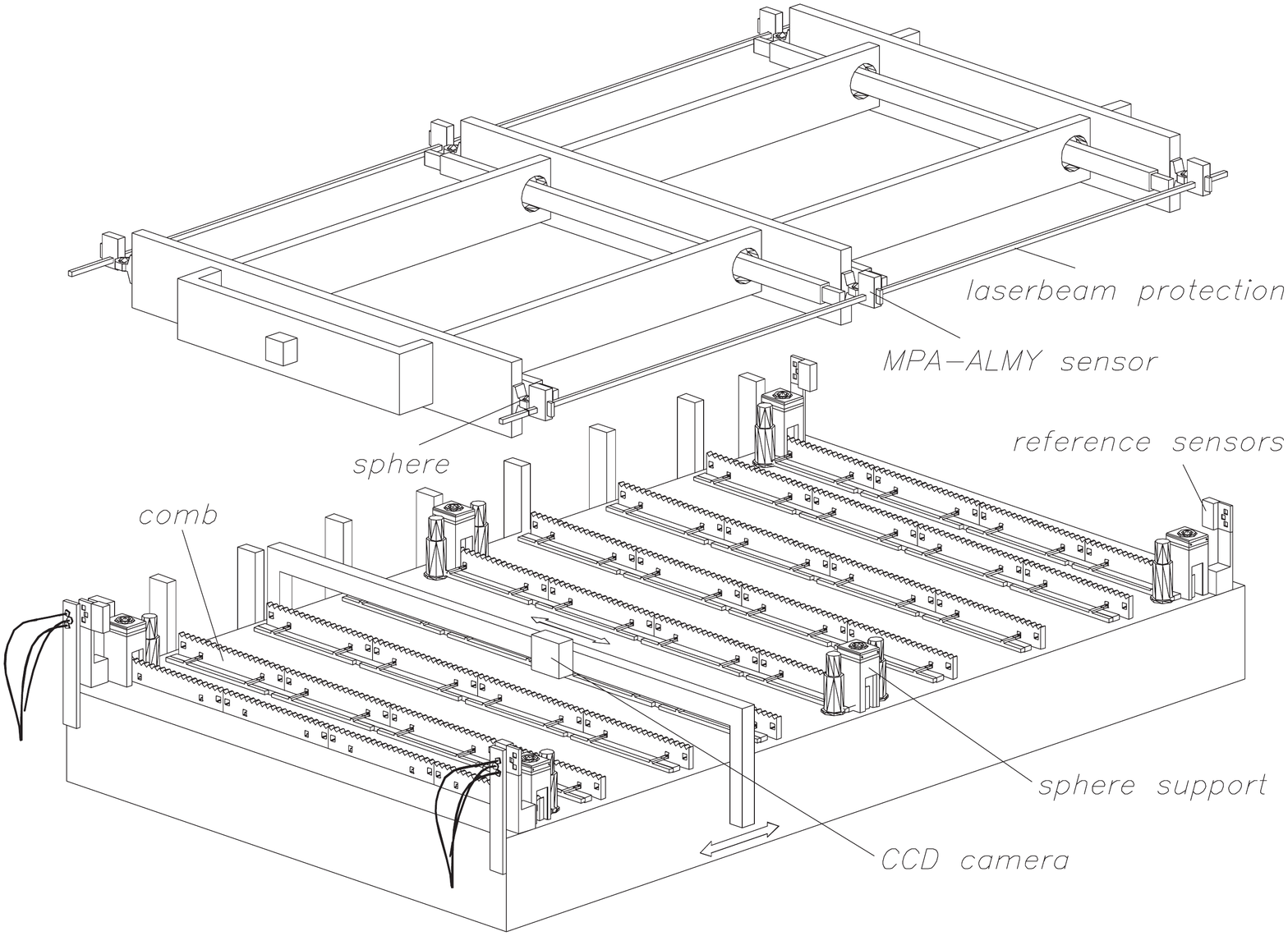}
\vspace{4mm}
\caption{MDT chamber assembly table\hfill\break
with MDT space frame and optical\hfill\break
monitoring devices.}\label{assembly}
\end{minipage} &
\begin{minipage}{0.48\textwidth}
\vspace{-19mm}
\includegraphics[width=1.02\textwidth]{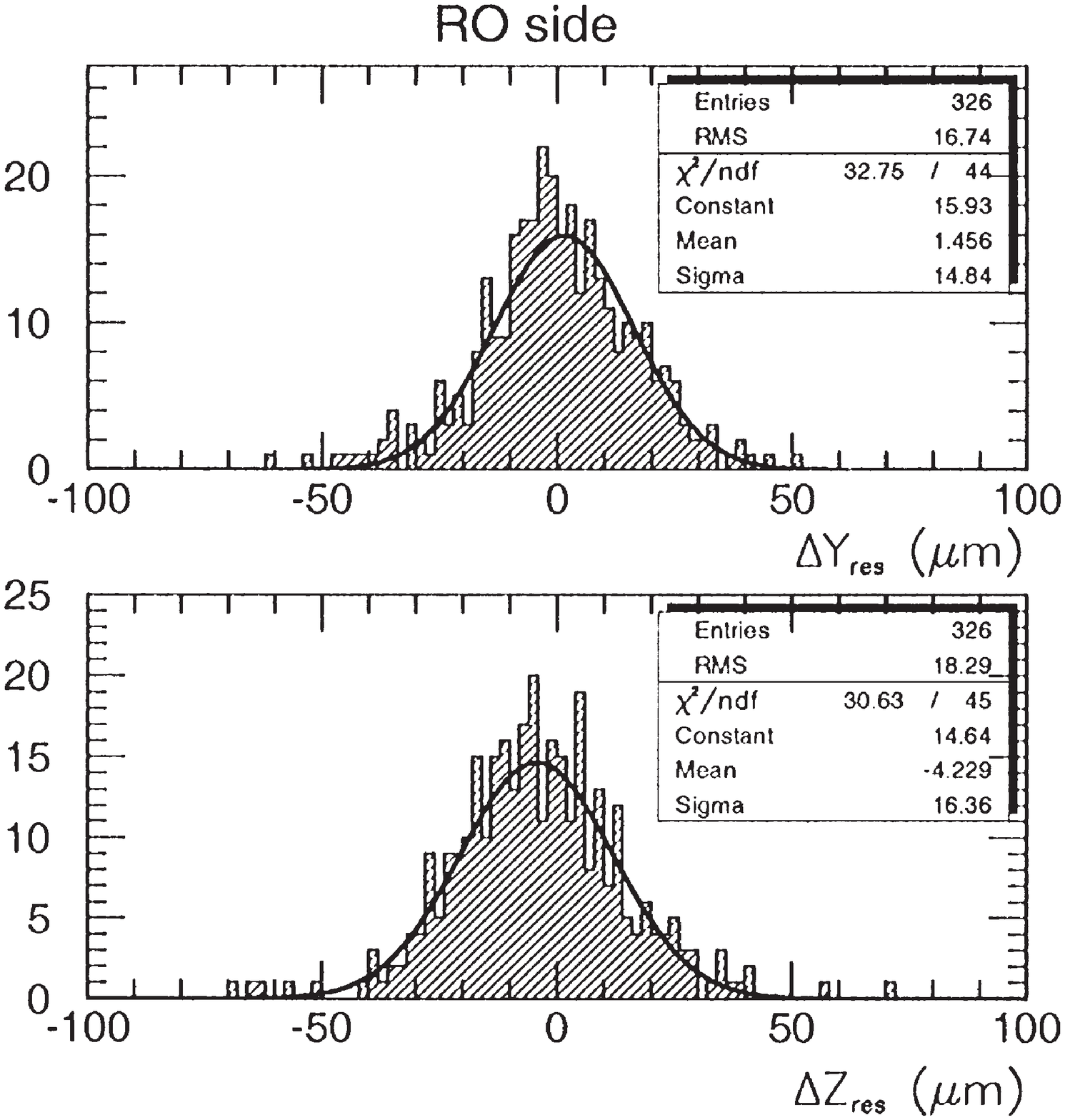}
\vspace{-20mm}
\caption{Residuals of measured wire\hfill\break
co\-ordi\-nates w.r.t.\ the expected grid.}\label{Xray}
\end{minipage} \\
\end{tabular}
\end{figure}

An improved version of the tube endplug designed for this method is shown
in Fig.~\ref{uvplug}. The wire is glued in a brass tube 
embedded in the insulating plastic (Noryl) body of the endplug to protect the wire 
against high voltage discharges. For long-term wire fixation, the wire is crimped in
a small copper tubelet at the end of the brass tube. 
Reliable ground contact to the aluminum tubes is provided by spot welds
to an aluminum ring on the endplug using a specially developed laser welding technique.
For the large-scale production, the wires are positioned
in the tubes using precisely machined endplugs. A wire positioning accuracy
of better than $9~\mu$m (rms) with respect to the endplug center has been achieved.
The drift tubes are assembled semi-automatically in a temperature-controlled
clean room. 

All drift tubes assembled in the chamber have to fulfil 
stringent quality criteria, including a gas leak rate at 3~bar
below $10^{-8}$~bar$\;\cdot\;$l/s, a
wire tension within $2\%$ of the nominal value and HV stability requirements.

For the assembly of the chamber~\cite{TDR}, 
the tubes of each layer are positioned with a precision of $4~\mu$m (rms) on combs
installed on a granite table in a climatized clean room 
(see Figs.~\ref{BOS} and \ref{assembly}).
Subsequent tube layers are glued to the space frame which 
is positioned above the combs with an accuracy of $\pm 5~\mu$m
monitored by optical measuring devices (see Fig.~\ref{assembly}).

The wire positions in the prototype chamber have been measured at CERN~\cite{Xtomo}
with an accuracy of about $5~\mu$m (rms) by scanning
the tube ends with stereo X-ray sources.
The distributions of the residuals of the measured wire coordinates y and z
with respect to the expected wire grid are shown in Fig.~\ref{Xray}.
A wire positioning accuracy of better than $18~\mu$m (rms)
has been achieved for one of the largest chambers in the ATLAS muon spectrometer.
\vspace{2mm}
\begin{figure}[h]
\begin{tabular}{ll}
\begin{minipage}{0.5\textwidth}
\hspace{-5mm}
\includegraphics[width=1.04\textwidth]{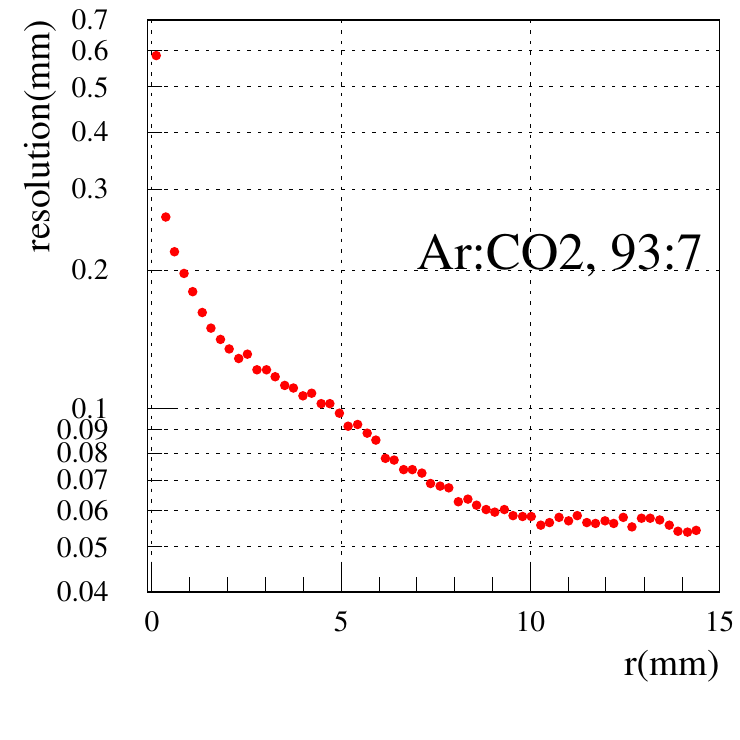}
\vspace{-6mm}
\caption{Single-tube resolution as a function of drift distance
for Ar:CO${}_2$ gas at 3 bar.}\label{resol}
\end{minipage} &
\begin{minipage}{0.5\textwidth}
\hspace{-5mm}
\includegraphics[width=1.04\textwidth]{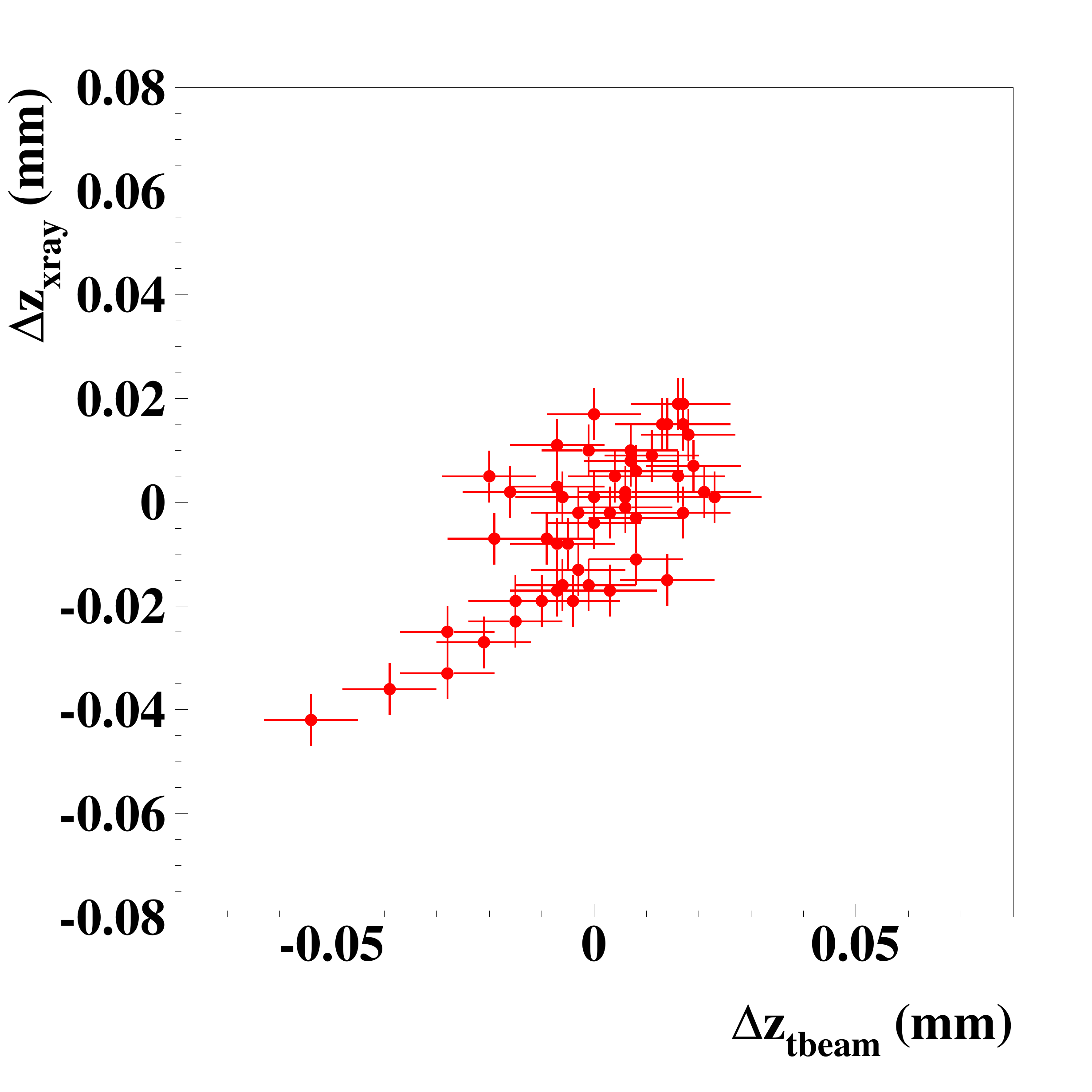}
\vspace{-2mm}
\caption{Correlation of wire coordinate\hfill\break
measurements with X-rays and with\hfill\break
tracks in the test beam.}\label{wirepos}
\end{minipage} \\
\end{tabular}
\end{figure}

The prototype chamber has been tested in a 300~GeV muon beam at CERN fully equipped
with readout electronics using as drift gas Ar:CO${}_2$ $(93:7)$ at 3~bar chosen
to prevent ageing of the drift tubes under LHC conditions. Using a silicon strip
detector telescope~\cite{silicon} as external reference, the space to drift-time relationship and the
position resolution as a function of the drift distance were determined (see Fig.~\ref{resol})
giving an average single-tube resolution of $70~\mu$m (rms). This r-t-relation
was applied to the other drift tubes in the beam adjusting only the maximum drift time.
Requiring the track residual distributions as function of the drift distance 
to be symmetric left and right of the wires 
provides information about displacements of the wires from their nominal positions.
Comparison with the X-ray measurements of the wire coordinates perpendicular
to the beam (z) shows a good correlation
(see Fig.~\ref{wirepos}) and agreement within $10~\mu$m (rms).
The tests showed that using Ar:CO${}_2$ with only $7\%$ CO${}_2$ as gas mixture 
requires great care in the fabrication and operation of the large number of drift tubes
in order to maintain high precision and reliability.

After the demonstration of the mechanical precision of the chamber construction
and experience with the chamber operation in the beam showing the required performance,
the production of MDT chambers for the ATLAS experiment has now started.


\end{document}